\documentclass{article}
\usepackage{daf99}
\begin{document}
\title{
CHIRAL DYMAMICS OF MANY PION SYSTEMS:\\ THE $\rho\to4\pi$ AND
$\omega\to5\pi$ DECAYS. }
\author{
N.~N.~Achasov        \\
{\em Lab. Theor. Physics, Sobolev Inst. for Mathematics} \\
A.~A.~Kozhevnikov        \\
{\em Lab. Theor. Physics, Sobolev Inst. for Mathematics}
}
\maketitle
\baselineskip=11.6pt
\begin{abstract}
Based on the Weinberg lagrangian or, in a new language, the lagrangian
of hidden local symmetry added with the term induced by the anomalous
lagrangian of Wess and Zumino,
the dynamics of the decays $\rho\to4\pi$ and $\omega\to5\pi$
is considered. The excitation curves of $\rho$ resonance in its decay to
$4\pi$ are obtained for the first time.
The comparison with recent CMD-2 data is given.
The partial widths of the decays $\omega\to2\pi^+2\pi^-\pi^0$ and
$\omega\to\pi^+\pi^-3\pi^0$ which can be measured on colliders with
luminosity $10^{33}\mbox{ cm }^{-2}\mbox{ s }^{-1}$ are evaluated for
the first time.
\end{abstract}
\baselineskip=14pt
\section{Introduction}
The decay  $\rho\to4\pi$ is a unique source of soft,
$|{\bf p}|\sim m_\pi$, pions, and can be used for the study of the chiral
dynamics of many-pion systems. By this reason it attracts much interest
\cite{rittenberg69},\cite{bramon93},\cite{kuraev95},\cite{birse96}.
It was found in papers \cite{rittenberg69,bramon93,kuraev95}
that the above decay should be rather strong,
B$(\rho\to4\pi)\sim10^{-4}$.
The calculations of Ref. \cite{bramon93,kuraev95}
 were analyzed in detail
in the work \cite{birse96},
where a number of shortcomings of
the former related with the actual violation of chiral invariance, in
particular, the Adler condition for soft pions, was uncovered,
The correct result obtained in \cite{birse96}, corresponds to
B$(\rho\to4\pi)\sim10^{-5}$.
The large magnitude of the branching ratio B$(\rho\to4\pi)\sim10^{-4}$
obtained in  \cite{rittenberg69} is related, in all appearance, with a
very rough method of calculation. A common drawback of all the above cited
papers is that their authors evaluate the partial width  at the only
energy equal to the mass of the $\rho$, as if the latter would be a
genuine narrow peak. However, the fact that the
width of the $\rho$ resonance is rather large and $\Gamma(\rho\to4\pi,E)$
rises rapidly with the energy  increase even at energies inside the
$\rho$ peak, push one to think that the
magnitude of the $4\pi$ partial width at the $\rho$ mass cannot be an
adequate characteristics of the dynamics of the process. In this respect,
it is just the resonance excitation curve in the channel $e^+e^-\to
\rho^0\to4\pi$ is of much interest, being a test ground of various
chiral models of the decay under consideration.

Here, based on the Weinberg lagrangian \cite{weinberg68}
obtained under the nonlinear realization of the chiral symmetry, or,
in modern terms, the lagrangian of hidden local symmetry (HLS)
\cite{bando}, the partial widths and resonance excitation curves are
calculated for the reactions $e^+e^-\to\rho^0\to2\pi^+2\pi^-$ and
$e^+e^-\to\rho^0\to\pi^+\pi^-2\pi^0$.
It is shown that the intensities of the above decays change as fast as two
times the phase space variation, upon the energy variation inside the
$\rho$ widths. All this means that $e^+e^-$ offers an ideal tool for the
study of such effects. The decay widths of charged $\rho$ meson,
$\rho^\pm\to\pi^\pm3\pi^0$ and  $\rho^\pm\to2\pi^\pm\pi^\mp\pi^0$,
as well as $\omega$ meson,
$\omega\to2\pi^+2\pi^-\pi^0$ and $\omega\to\pi^+\pi^-3\pi^0$, are also
evaluated.

\section{The $\rho\to4\pi$ decay amplitudes}
The HLS approach \cite{bando} permits one to include the axial
mesons as well. An ideal treatment would consist of that under the
assumption of $m_\rho\sim E\ll m_{a_1}$, the difference between
the models with and without $a_1$ meson would be reduced to the
taking into account the higher derivatives and would be small. In
real life one has $m^2_{a_1}-m^2_\rho\sim m^2_\rho$, and the
correction may appear to be appreciable even at the $\rho$ mass.
In fact, the calculation \cite{birse96} shows that the corrections
amounts to $\sim20\mbox{ - }30\%$ in the width. This means, in
particular, that the left shoulder of the $\rho$ peak, where the
contributions of higher derivatives are vanishing rapidly, is the
best place to work. We do not take into account $a_1$ meson in the
present work.

The amplitudes of the decays of our interest are obtained from the
Weinberg lagrangian \cite{weinberg68}
\begin{eqnarray}
{\cal L}&=&-{1\over4}\left(\partial_\mu\mbox{\boldmath$\rho$}_\nu-
\partial_\nu\mbox{\boldmath$\rho$}_\mu+
g_\rho[\mbox{\boldmath$\rho$}_\mu\times\mbox{\boldmath$\rho$}_\nu]
\right)^2 +{m^2_\rho\over2}\left\{\mbox{\boldmath$\rho$}_\mu+
{[\mbox{\boldmath$\pi$}\times\partial_\mu\mbox{\boldmath$\pi$}]
\over 2g_\rho
f^2_\pi\left(1+{\mbox{\boldmath$\pi$}^2\over4f^2_\pi}
\right)}\right\}^2 \nonumber\\
&&+{(\partial_\mu\mbox{\boldmath$\pi$})^2
\over2(1+\mbox{\boldmath$\pi$}^2/4f^2_\pi)^2}
-{m^2_\pi\mbox{\boldmath$\pi$}^2\over2\left(1+\mbox{\boldmath$\pi$}^2
/4f^2_\pi\right)}, \label{lwein}
\end{eqnarray}
where $f_\pi=92.4$ MeV is the pion decay constant, and
$g_\rho=g_{\rho\pi\pi}$, if one demands the universality condition
. Then the so called  KSRF relation \cite{ksrf}
$2g^2_{\rho\pi\pi}f^2_\pi/m^2_\rho=1$  arises, which beautifully
agrees with experiment. The $\rho\pi\pi$ coupling constant
resulting from this relation is $g_{\rho\pi\pi}=5.89$. Introducing
the 4-vector of polarization of the decaying $\rho$ meson,
$\varepsilon_\mu$, one can write the amplitudes in the form
$$M={g_{\rho\pi\pi}\over f^2_\pi}\varepsilon_\mu J_\mu.$$ Let us
give the expressions for the current $J_\mu$ for all the decay
modes considered here.\\ 1) The decay
$\rho^0(q)\to\pi^+(q_1)\pi^+(q_2)\pi^-(q_3)\pi^-(q_4)$. One has
\begin{eqnarray}
J_\mu&=&(1+P_{12})(1+P_{34})\left\{-q_{1\mu}\left[{1\over2}
+{2(q_2,q_3)\over D_\pi(q-q_1)}\right]+q_{3\mu}\left[{1\over2}
+{2(q_1,q_4)\over D_\pi(q-q_3)}\right]\right.   \nonumber\\
&&\left.+(2g_{\rho\pi\pi}f_\pi)^2(1+P_{13})
{q_{1\mu}(q_3,q_2-q_4)\over D_\pi(q-q_1)D_\rho(q_2+q_4)}\right\}.
\label{eech}
\end{eqnarray}
Hereafter  $D_\pi(q)=m^2_\pi-q^2$ and $D_\rho(q)=m^2_\rho-q^2$ are inverse
propagators of pion and  $\rho$ meson, respectively, and $P_{ij}$
is the operator of interchange of the pion momenta $q_i$ and $q_j$.
Direct numerical evaluation shows that the neglect of the $\rho$ width is
justifiable with accuracy better than  $5\%$
up to the energy  $\sqrt{s}\leq0.9$ GeV. Hereafter $s$ stands for the
total center-of-mass energy squared. Recall that the allowing for the
finite widths effects is in fact equivalent to the loop correction being
taken into account.\\
2) The decay $\rho^0(q)\to\pi^+(q_1)\pi^-(q_2)\pi^0(q_3)\pi^0(q_4)$.
In this case one has $J_\mu=J^{\rm nan}_\mu+J^{\rm an}_\mu$, where
\begin{eqnarray}
J^{\rm nan}_\mu&=&-(1-P_{12})(1+P_{34})q_{1\mu}\left\{{1\over4}+
{1\over D_\pi(q-q_1)}\left[(q_3,q_4)\right.\right.   \nonumber\\
&&\left.\left.+(2g_{\rho\pi\pi}f_\pi)^2 {(q_3,q_2-q_4)\over
D_\rho(q_2+q_4)}\right]\right\}
+(1+P_{34}){(g_{\rho\pi\pi}f_\pi)^2\over
D_\rho(q_1+q_3)D_\rho(q_2+q_4)} \nonumber\\
&&\times\left[(q_1+q_3-q_2-q_4)_\mu(q_1-q_3,q_2-q_4)\right.\nonumber\\
&&\left.-2(q_1-q_3)_\mu(q_1+q_3,q_2-q_4)+
2(q_2-q_4)_\mu(q_2+q_4,q_1-q_3)\right] \label{eenena}
\end{eqnarray}
is obtained from Eq. (\ref{lwein}), while the contribution of the term
induced by the anomalous lagrangian of Wess and Zumino
\cite{bando,wza},
\begin{equation}
{\cal L}_{\omega\rho\pi}={N_cg^2_{\rho\pi\pi}\over8\pi^2 f_\pi}
\varepsilon_{\mu\nu\lambda\sigma}\partial_\mu\omega_\nu
\left(\mbox{\boldmath$\pi$}
\cdot\partial_\lambda\mbox{\boldmath$\rho$}_\sigma\right),
\label{wz}
\end{equation}
manifesting in the process  $\rho^0\to\omega\pi^0\to
\pi^+\pi^-\pi^0\pi^0$, is given by the expression
\begin{eqnarray}
J^{\rm an}_\mu&=&2\left({N_cg^2_{\rho\pi\pi}\over8\pi^2}\right)^2
(1+P_{34})\left[q_{1\mu}(1-P_{23})(q,q_2)(q_3,q_4)\right.\nonumber\\
&&\left.+q_{2\mu}(1-P_{13})(q,q_3)(q_1,q_4)+
q_{3\mu}(1-P_{12})(q,q_1)(q_2,q_4)\right]        \nonumber\\
&&\times\left[{1\over D_\rho(q_1+q_2)}+{1\over D_\rho(q_1+q_3)}
+{1\over D_\rho(q_2+q_3)}\right]{1\over D_\omega(q-q_4)},
\label{eenean}
\end{eqnarray}
where $D_\omega(q)=m^2_\omega-q^2$ is the inverse $\omega$ meson
propagator, and  $N_c=3$ is the number of colors. In general, this
term is attributed to the contribution of higher derivatives.
Nevertheless, we take it into account to show the effect of the
latter and the dynamical effect of the opening of the channel
$\rho\to\omega\pi\to4\pi$. In agreement with \cite{bando}, the
contribution of the point vertex $\omega\to3\pi$ is omitted. The
following amplitudes of the charged $\rho$ decay are necessary for
obtaining the $\omega\to5\pi$ decay amplitude, and are of their
own interest when studying the reactions of peripheral $\rho$
meson production and the $\tau\to\nu_\tau4\pi$ decay. \\ 3) The
decay $\rho^+(q)\to\pi^+(q_1)\pi^0(q_2)\pi^0(q_3)\pi^0(q_4)$. One
has
\begin{eqnarray}
J_\mu&=&2q_{1\mu}\left[1+{(q_2,q_3)+(q_2,q_4)+(q_3,q_4)\over
D_\pi(q-q_1)}\right]-(1+P_{23}){2q_{2\mu}(q_3,q_4)\over
D_\pi(q-q_2)} \nonumber\\ &&-{2q_{4\mu}(q_2,q_3)\over
D_\pi(q-q_4)} -(2g_{\rho\pi\pi}f_\pi)^2(1+P_{23})\left[(1+P_{34})
\right.   \nonumber\\
&&\left.\times{q_{2\mu}(q_4,q_1-q_3)\over
D_\pi(q-q_2)D_\rho(q_1+q_3)} +{q_{4\mu}(q_2,q_1-q_3)\over
D_\pi(q-q_4)D_\rho(q_1+q_3)}\right] \label{peneu}
\end{eqnarray}
4) The decay $\rho^+(q)\to\pi^+(q_1)\pi^+(q_2)\pi^-(q_3)\pi^0(q_4)$.
Here, the contribution induced by the anomalous lagrangian of
Wess and Zumino  is also possible, hence
$J_\mu=J^{\rm nan}_\mu+J^{\rm an}_\mu$,  where
\begin{eqnarray}
J^{\rm nan}_\mu&=&(1+P_{12})\left\{(1-P_{14})q_{1\mu}\left[{1\over2}
+{2(q_2,q_3)\over D_\pi(q-q_1)}\right]-(2g_{\rho\pi\pi}f_\pi)^2
\left[(1+P_{23})\right.\right.     \nonumber\\
&&\left.\left.\times{q_{1\mu}(q_2,q_3-q_4)\over
D_\pi(q-q_1)D_\rho(q_3+q_4)}+{q_{4\mu}(q_1,q_2-q_3)\over
D_\pi(q-q_4)D_\rho(q_2+q_3)}\right]\right\}
\nonumber\\
&&+(g_{\rho\pi\pi}f_\pi)^2(1+P_{12})
\left\{\left[(q_1+q_3-q_2-q_4)_\mu(q_1-q_3,q_2-q_4)
\right.\right.\nonumber\\
&&\left.\left.-2(q_1-q_3)_\mu(q_1+q_3,q_2-q_4)
+2(q_2-q_4)_\mu(q_1-q_3,q_2+q_4)\right]\right.\nonumber\\
&&\left.\times{1\over D_\rho(q_1+q_3)D_\rho(q_2+q_4)}\right\}
\label{pechnan}
\end{eqnarray}
is obtained from Eq.~(\ref{lwein}), while the term induced by the
anomaly looks as
\begin{eqnarray}
J^{\rm an}_\mu&=&2\left({N_cg^2_{\rho\pi\pi}\over8\pi^2}\right)^2
(1+P_{23})\left[q_{1\mu}(1-P_{24})(q,q_4)(q_2,q_4)\right.\nonumber\\
&&\left.+q_{2\mu}(1-P_{14})(q,q_1)(q_3,q_4)+
q_{4\mu}(1-P_{12})(q,q_2)(q_1,q_3)\right]        \nonumber\\
&&\times\left[{1\over D_\rho(q_1+q_2)}+{1\over D_\rho(q_1+q_4)}
+{1\over D_\rho(q_2+q_4)}\right]{1\over D_\omega(q-q_3)}.
\label{pechan}
\end{eqnarray}
One can verify that up to the corrections of the order of
$\sim\mu^2/m^2_\rho$, the above written amplitudes vanish in the limit
of vanishing 4-momentum of each final pion. In other words, they
obey the Adler condition.

\section{Calculating the $\rho\to4\pi$ decay widths and
excitation curves.}

 The results of evaluation of the partial widths at
$\sqrt{s}=m_\rho=770$ MeV are as follows:
$\Gamma(\rho^0\to2\pi^+2\pi^-,m^2_\rho)=0.89$ keV,
$\Gamma(\rho^0\to\pi^+\pi^-2\pi^0,m^2_\rho)=0.24$ keV and 0.44 keV,
respectively, without and with the induced anomaly term being taken into
account. This coincides with the results obtained in \cite{birse96}.
In the case of the charged $\rho$ meson decays it is obtained for
the first time:
$\Gamma(\rho^+\to\pi^+3\pi^0,m^2_\rho)=0.41$ keV,
$\Gamma(\rho^+\to2\pi^+\pi^-\pi^0,m^2_\rho)=0.71$ keV and 0.90 keV
respectively, without and with the anomaly induced term being taken into
account.

\begin{figure}[t]
 \vspace{9.0cm}
\includegraphics{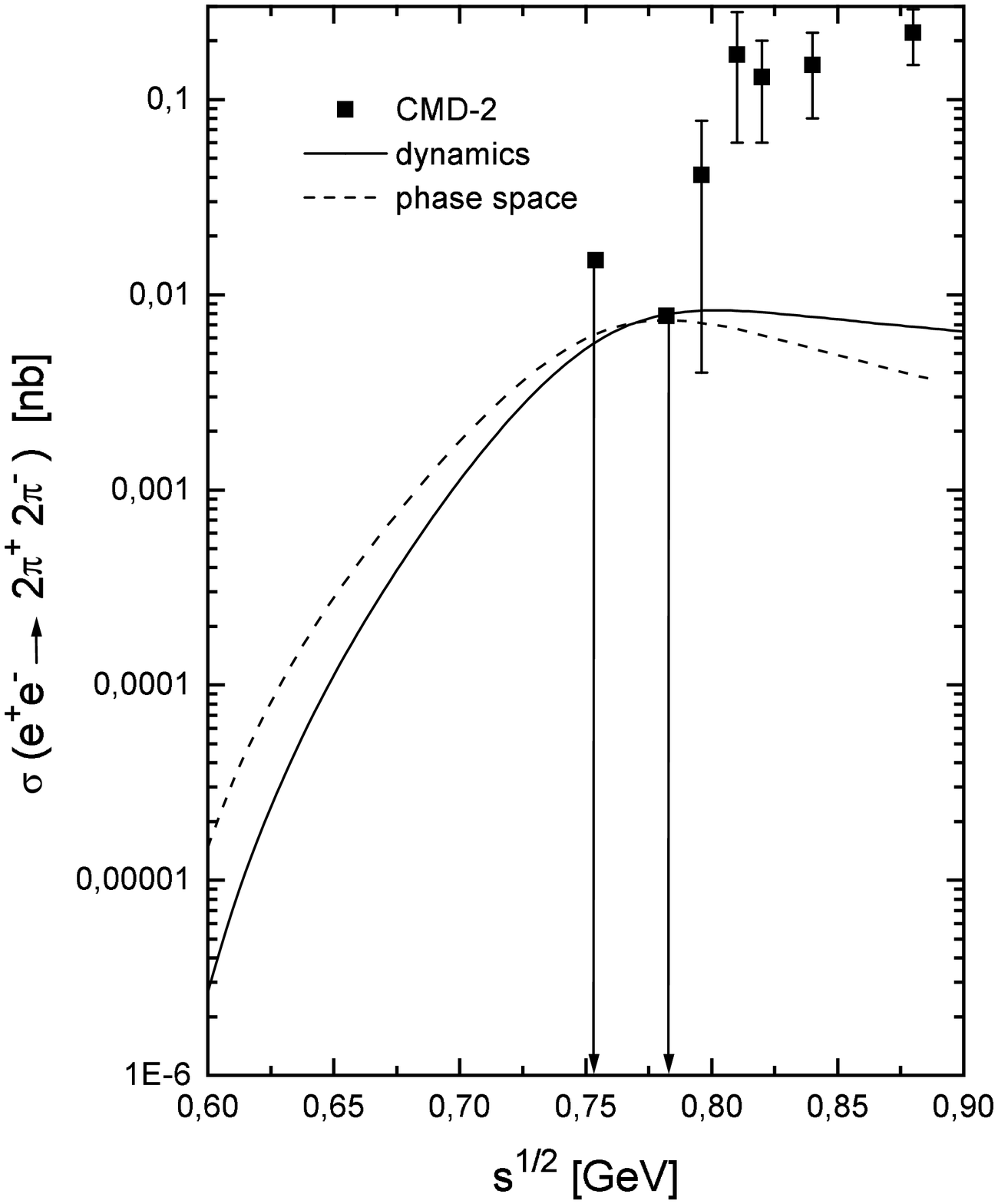}
 \caption{\it
The energy dependence of the
$e^+e^-\to\rho^0\to\pi^+\pi^-\pi^+\pi^-$ reaction cross section in the
model based on the chiral lagrangian due to  Weinberg,
Experimental points are from \protect\cite{cmd2}.
    \label{fig1} }
\end{figure}

\begin{figure}[t]
 \vspace{9.0cm}
\includegraphics{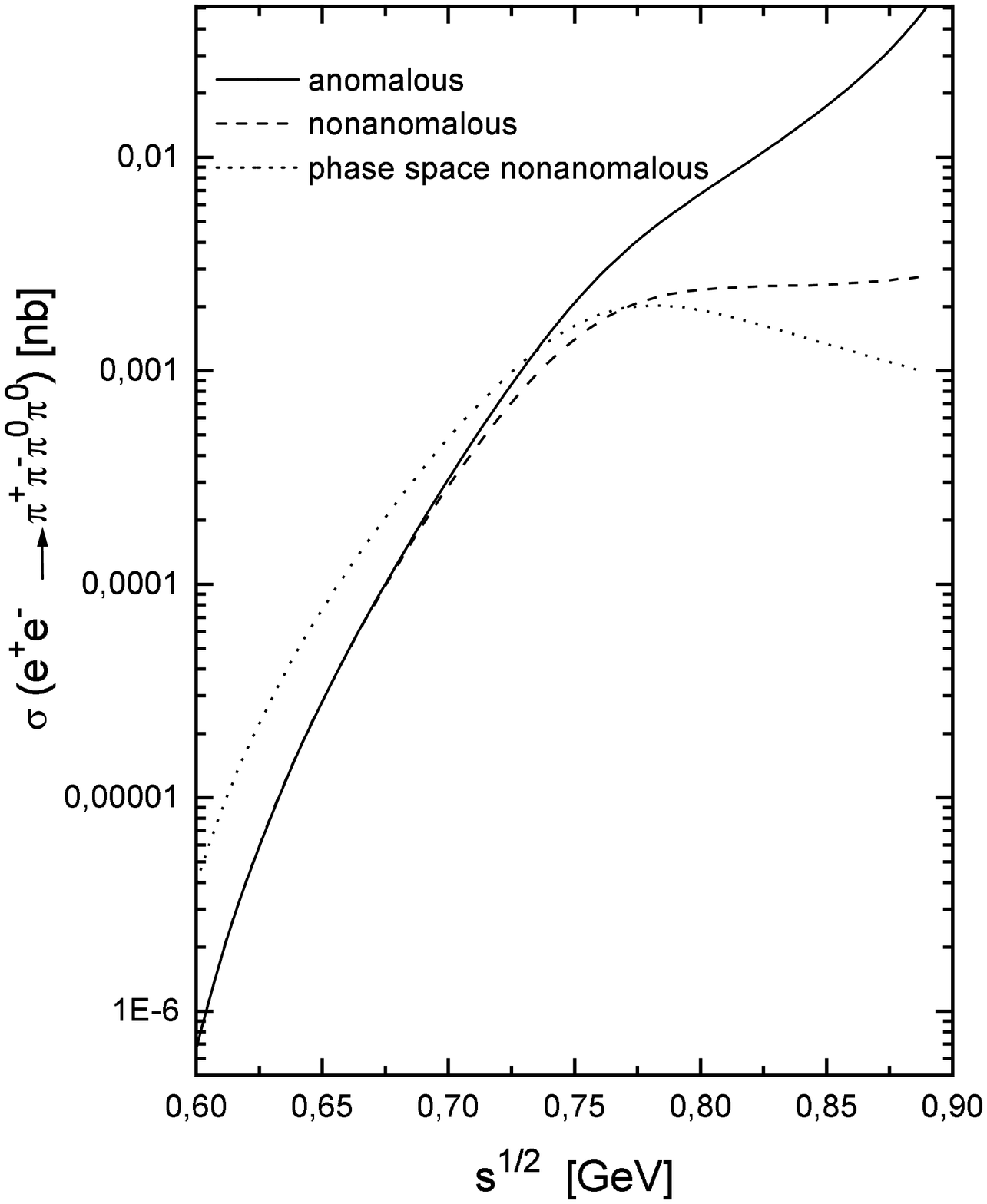}
 \caption{\it
The energy dependence of the
$e^+e^-\to\rho^0\to\pi^+\pi^-\pi^0\pi^0$ reaction cross section in the
model based on the chiral lagrangian due to  Weinberg.
    \label{fig2} }
\end{figure}

The results of the $4\pi$ state production cross section in the
reaction $e^+e^-\to\rho^0\to4\pi$ are shown in Fig.~\ref{fig1} and
\ref{fig2}. To demonstrate the effects of chiral dynamics, also
shown is the energy dependence of the cross section, which
expression contains the quantity $\Gamma^{\rm LIPS}(s)$
representing the width evaluated in the model of the simple $4\pi$
phase space normalized to the widths at the $\rho$ mass. When so
doing, the ratio $\Gamma(\rho\to2\pi^+2\pi^-,s)/\Gamma^{\rm
LIPS}(\rho\to2\pi^+2\pi^-,s)$ changes from 0.4 at $\sqrt{s}=650$
MeV to 1 at $\sqrt{s}=m_\rho$. As can be observed from the
figures, the rise of the $\rho\to4\pi$ partial width with the
energy increase is such fast that it compensates completely the
falling of the $\rho$ meson propagator and electronic width. Also
noticeable is the dynamical effect in the decay
$\rho^0\to\omega\pi^0\to\pi^+\pi^-2\pi^0$ at $\sqrt{s}>850$ MeV
resulting from the anomaly induced lagrangian $\omega\pi$
threshold. See Fig.~\ref{fig2}. To quantify the above mentioned
effect of vanishing contribution of higher derivatives at the left
shoulder of the $\rho$ resonance it should be noted that the
difference between magnitudes of
$\Gamma(\rho\to\pi^+\pi^-2\pi^0,s)$ with and without term
originating from the anomaly induced lagrangian, equal to $100\%$
at $\sqrt{s}=m_\rho$, diminishes rapidly with the energy decrease
amounting to $8\%$ at  $\sqrt{s}=700$ MeV and $0.3\%$ at
$\sqrt{s}=650$ MeV.

As it is seen from Fig.~\ref{fig1}, the predictions of chiral symmetry
for the $e^+e^-\to2\pi^+2\pi^-$ reaction cross section do not contradict
to the three lowest experimental points of CMD-2 detector \cite{cmd2}.
However, at $\sqrt{s}>800$ MeV one can observe a substantial deviation
between the predictions of the lagrangian (\ref{lwein}) the data of
CMD-2. In all appearance, this is due to the contribution of higher
derivatives and chiral loops neglected in the present work.
It is expected that the left shoulder of the $\rho$ is practically free
of such contributions, and by this reason it is preferable for
studying the dynamical effects of chiral symmetry. Note that even at
$\sqrt{s}=650$ MeV, where the contribution of higher derivatives is
negligible, one can hope to gather one event of the reaction
$e^+e^-\to2\pi^+2\pi^-$, provided the luminosity
$L=10^{32}\mbox{cm}^{-2}\mbox{s}^{-1}$ is achieved, and up to 10
events per day at $\sqrt{s}=700$ MeV, i.e. to have a factory
for a comprehensive study of the chiral dynamics of many-pion systems.

\section{The $\omega\to5\pi$ decay widths.}

One may convince oneself that the $\omega\to\rho\pi\to5\pi$ decay
amplitude unambiguously results from the anomaly induced lagrangian
(\ref{wz}). In order to calculate it, notice that, although
$|{\bf q}_\pi|/m_\pi\simeq0.5$,
the nonrelativistic expressions describe the
$\rho\to4\pi$ decay widths with the accuracy $10\%$ in the $4\pi$
mass range relevant for the present purpose. Using the corresponding
limiting expressions for currents (\ref{eech}), (\ref{eenena}),
(\ref{peneu}), (\ref{pechnan}), and neglecting the corrections of the
order of $O(|{\bf q}_\pi|^4/m^4_\pi)$ or higher, one obtains
\begin{eqnarray}
M(\omega\to2\pi^+2\pi^-\pi^0)&=&{N_c\over(2\pi)^2}\left({g_{\rho\pi\pi}
\over2f_\pi}\right)^3\varepsilon_{\mu\nu\lambda\sigma}q_\mu\epsilon_\nu
\left[(1+P_{12}){q_{1\lambda}(q_2+3q_4)_\sigma\over D_\rho(q-q_1)}
\right.\nonumber\\
&&\left.-(1+P_{35}){q_{3\lambda}(q_5+3q_4)_\sigma\over D_\rho(q-q_3)}
+4{q_{4\lambda}(q_3+q_5)_\sigma\over D_\rho(q-q_4)}\right],
\label{om3pi0}
\end{eqnarray}
with the final momentum  assignment according to $\pi^+(q_1)$,
$\pi^+(q_2)$, $\pi^-(q_3)$, $\pi^-(q_5)$, $\pi^0(q_4)$; and
\begin{eqnarray}
M(\omega\to\pi^+\pi^-3\pi^0)&=&{N_c\over(2\pi)^2}\left({g_{\rho\pi\pi}
\over2f_\pi}\right)^3\varepsilon_{\mu\nu\lambda\sigma}q_\mu\epsilon_\nu
\left\{(1-P_{12}){4q_{2\lambda}q_{1\sigma}\over D_\rho(q-q_1)}
\right.       \nonumber\\
&&\left.+(q_1-q_2)_\lambda\left[{q_{3\sigma}\over D_\rho(q-q_3)}
+{q_{4\sigma}\over D_\rho(q-q_4)}\right.\right.\nonumber\\
&&\left.\left.+{q_{5\sigma}\over D_\rho(q-q_5)},
\right]\right\}.
\label{om1pi0}
\end{eqnarray}
with the final momentum  assignment according to $\pi^+(q_1)$,
$\pi^-(q_2)$, $\pi^0(q_3)$, $\pi^0(q_4)$, $\pi^0(q_5)$. In both
above formulas, $\epsilon_\nu$, $q_\mu$ stand for four-vectors of
polarization and momentum of $\omega$ meson.

Since the invariant mass of the $4\pi$ system changes in
very narrow range $558\mbox{ MeV}<m_{4\pi}<642\mbox{ MeV}$,
one can set it in all the $\rho$ propagators,
with the accuracy $20\%$ in width,
to the equilibrium value $\bar m_{4\pi}=620$ MeV evaluated
for the pion energy $E_\pi=m_\omega/5$ which gives the dominant
contribution.
The resulting $\omega\to2\pi^+2\pi^-\pi^0$ and
$\omega\to\pi^+\pi^-3\pi^0$ decay amplitudes equal to
5/2 times the
$\omega\to\rho^0\pi^0\to2\pi^+2\pi^-\pi^0$ and
$\omega\to\rho^+\pi^-\to\pi^+\pi^-3\pi^0$ decay amplitudes, respectively.

The evaluation of the partial widths valid with accuracy $20\%$
can be obtained upon taking the amplitude of each considered decays as
5/2 times the $\rho\pi$ production state amplitude with the subsequent
decay $\rho\to4\pi$, and calculate the partial width
using the calculated widths of the latter:
\begin{eqnarray}
B_{\omega\to2\pi^+2\pi^-\pi^0}&=&\left({5\over2}\right)^2
{2\over\pi\Gamma_\omega}
\int_{4m_{\pi^+}}^{m_\omega-m_{\pi^0}}m^2\Gamma_{\omega\to\rho^0\pi^0}(m)
  \nonumber\\
&&\times{\Gamma_{\rho\to2\pi^+2\pi^-}(m)\over|D_\rho(m^2)|^2}dm
=1.6\times10^{-9}; \label{b1pi0}
\end{eqnarray}
где $\Gamma_{\omega\to\rho^0\pi^0}(m)=g^2_{\omega\rho\pi}q^3
(m_\omega,m,m_{\pi^0})/12\pi$,
$g_{\omega\rho\pi}=N_cg^2_{\rho\pi\pi}/8\pi^2f_\pi=14.3$ GeV$^{-1}$,
and  $q(m_a,m_b,m_c)$ is the momentum of final particle $b$ (or $c$)
in the rest frame system of decaying particle $a$. The partial width
$B_{\omega\to\pi^+\pi^-3\pi^0}=1.2\times10^{-9}$
is obtained from Eq.~(\ref{b1pi0}) upon inserting the lower integration
limit to $m_{\pi^+}+3m_{\pi^0}$, the upper one to
$m_\omega-m_{\pi^0}$, and $m_{\pi^0}\to m_{\pi^+}$
in the expression for the momentum $q$ and substitution
of the $\rho^+\to\pi^+3\pi^0$ decay width corrected for the mass
difference of charged and neutral pions. Provided the luminosity
$L=10^{33}\mbox{cm}^{-2}\mbox{s}^{-1}$
is achieved, one may expect about 3 events per week
for the considered decays.

\section{Acknowledgements.}

We are grateful to A.~E.~Bondar, U.~-G.~Meissner, G.~N.~Shestakov
and A.~I.~Sukhanov for discussions. Special thanks to A.~E.~B. and
A.~I.~S. of the above who kindly supplied us with the table of
experimental data. The present work is supported in part by the
grant RFBR-INTAS IR-97-232.


\end{document}